# Theoretical investigations of the origin of persistent luminescence in spinel oxides $MgGa_2O_4$ and $MgAl_2O_4$

Xiuli Yang[1,2], Ran Zhou[1], Hongliang Shi[1,*], Yifeng Duan[3,*], and Mao-Hua Du[4]

[1]Department of Physics, Beihang University, Beijing 100191, China

[2]Department of physics, Jinan University, Jinan, Shandong Province, China

[3]School of Materials and Physics, China University of Mining and Technology, Xuzhou, Jiangsu 221116, China

[4]NAGA, Knoxville, TN 37934, USA

Keywords: oxides, defects, luminescence, radiative, nonradiative, DFT calculation

*Corresponding Author: Hongliang Shi (hlshi@buaa.edu.cn), Yifeng Duan (yifeng@cumt.edu.cn)

# Abstract

MgGa$_2$O$_4$ and MgAl$_2$O$_4$ have attracted significant interest due to their unique intrinsic persistent luminescence, offering promising potential for various applications. In this paper, from the perspective of defect physics, we systemically investigate the origin of persistent luminescence phenomena in pristine MgGa$_2$O$_4$ and MgAl$_2$O$_4$, employing accurate hybrid functional calculations. Our results show that vacancies and antisite defects involving the two cations are the dominant point defects in both materials. Our calculated optical excitation and emission peaks associated with the Mg$_{Ga}$ defect agree well with the experimentally observed blue luminescence peak at about 2.9 eV in MgGa$_2$O$_4$. In MgAl$_2$O$_4$, the intradefect optical transition within the V$_O$-Mg$_{Al}$ donor-acceptor defect complex is identified as a likely origin for the observed 2.7 eV emission peak. Furthermore, the calculated radiative recombination coefficients of Mg$_{Ga}$ and V$_O$-Mg$_{Al}$ are significantly higher than their nonradiative counterparts, supporting their roles as efficient luminescent centers. Our results regarding the optical processes of oxygen vacancy V$_O$ in MgGa$_2$O$_4$ and MgAl$_2$O$_4$ are also in good agreement with experimental results. Based on the calculated defect thermodynamic transition levels, the intrinsic persistent luminescence in MgGa$_2$O$_4$ and MgAl$_2$O$_4$ may be attributed to electron traps, Ga$_{Mg}$ and V$_O$, in the former and a hole trap, Mg$_{Al}$, in the latter. Donor-acceptor defect complexes (V$_O$+V$_{Mg}$ and V$_O$+V$_{Ga}$) are also found to serve as effective carrier trapping centers in MgGa$_2$O$_4$. The calculated trap depths are also consistent with thermoluminescence spectroscopy measurements.

# I. Introduction

Highly efficient persistent luminescence materials, such as $SrAl_2O_4:Eu^{2+},Dy^{3+}$ and $Y_2O_3:Eu^{2+}$, have found important applications in areas including in vivo imaging [1] and safety and emergency signage [2]. Lanthanide or transition metal ions are typically doped into these materials, acting as emission centers. The delayed luminescence is generally attributed to a trapping-detrapping process involving trap centers [1, 2].

Recent studies showed that undoped ternary compound oxides $MgGa_2O_4$ and $MgAl_2O_4$ also exhibit stable and broad persistent emissions under UV excitation [3-8]. Their excellent photoluminescence properties enable a wide range of potential applications. In contrast to conventional materials like $SrAl_2O_4$ and $Y_2O_3$, $MgGa_2O_4$ and $MgAl_2O_4$ exhibit self-activated persistent luminescence, meaning that no extrinsic dopants are required [3-8]. In such cases, both luminescent centers and traps, which causes delayed luminescence, originate solely from intrinsic defects.

While most persistent luminescence materials rely on activator doping, self-activated ones have also been reported [6, 8]. Although the general mechanism underlying the persistent luminescence is understood, the detailed processes in specific materials, such as pristine $MgGa_2O_4$ and $MgAl_2O_4$, are still under debate [3-9]. Investigating self-activated $MgGa_2O_4$ and $MgAl_2O_4$ may provide new insights into persistent luminescence mechanisms.

Both $MgGa_2O_4$ and $MgAl_2O_4$ adopt a spinel structure, in which Mg occupies tetrahedral sites and Ga (or Al) occupies octahedral sites. Studies have shown that antisite defects, where Mg and Ga(Al) exchange positions, occur readily, and in some $AB_2O_4$ spinels, even an inverse spinel structure can form [6, 10-12]. In addition to antisites ($Mg_{Ga(Al)}$, $Ga(Al)_{Mg}$), vacancies ($V_O$ and $V_{Mg}$, $V_{Ga(Al)}$) are also present in $MgGa_2O_4$ and $MgAl_2O_4$. Moreover, defect complexes can form between these point defects. This variety of defects introduce significant complexity, and the microscopic origin of persistent luminescence in MgGa$_2$O$_4$ and MgAl$_2$O$_4$ remain under debate [3-8, 13-21].

In MgGa$_2$O$_4$, synthesized via high-temperature solid state reactions, bright blue luminescence was observed in the range of 350-600 nm with a peak at 415 nm (2.99 eV) and a full width at half maximum (FWHM) of 113 nm (0.84 eV). The donor–acceptor pair V$_O$-V$_{Ga}$ has been proposed as the recombination center responsible for this emission [3]. Another group reported a broad emission band in the range of 325-650 nm with a peak at 424 nm (2.92 eV) and a secondary peak at 480 nm (2.58 eV) [6]. They attributed these emission peaks to donor-acceptor pairs involving V$_O$ and V$_{Mg}$/V$_{Ga}$ [6]. In MgGa$_2$O$_4$ prepared by solution combustion synthesis, a broad blue emission from 350 to 600 nm with a peak at 450 nm (2.76 eV) was observed; it was attributed to oxygen vacancies within GaO$_6$ octahedron with no involvement from antistites and cation vacancies [15].

In MgAl$_2$O$_4$, a luminescence band with its peak at 2.69 eV was reported in thermochemically reduced samples and was attributed to the F center (V$_O$) [4]. A similar emission peak at 2.7 eV was also observed in samples obtained by high-temperature heating of powder mixtures and was assigned to the F$^+$ ( V$_O^+$ ) center [7]. Another emission peak at 2.36 eV was found in MgAl$_2$O$_4$ synthesized via gel combustion, with oxygen-related vacancies proposed as the luminescent centers [21]. Considering the coexistence of numerous antisite defects and vacancies in both MgGa$_2$O$_4$ and MgAl$_2$O$_4$ [3, 4, 6, 7], the exact roles of these intrinsic defects and their complexes in emission properties remain unresolved and merit further investigation, particularly due to their implications for defect physics and material applications.

In this work, we investigate the intrinsic defects and their impact on the luminescent properties of MgGa$_2$O$_4$ and MgAl$_2$O$_4$ using the hybrid density functional theory (DFT) calculations. Our results show that cation antisites, oxygen vacancies, and their donor-acceptor complexes in these materials strongly affect their self-activated luminescence properties. The most likely luminescent centers are identified based on our calculated excitation and emission energies, along with radiative and nonradiative recombination coefficients. Our study further suggests that electron-trapping centers such as $Ga_{Mg}$ and $V_O$ (isolated or within complexes $V_O+V_{Mg}$ and $V_O+V_{Ga}$) in $MgGa_2O_4$, as well as hole-trapping centers such as $Mg_{Al}$ in $MgAl_2O_4$, play key roles in the trapping-detrapping processes underlying persistent luminescence. This comprehensive and systematic study provides fundamental understanding of persistent luminescence in $MgGa_2O_4$ and $MgAl_2O_4$ and pave the way for future design of new materials with improved properties.

## II. Methods

Our theoretical approach is based on density functional theory (DFT), implemented using the Vienna Ab initio Simulation Package (VASP) [22]. To analyze the defects and their optical properties in $MgGa_2O_4$ and $MgAl_2O_4$, we employed the Heyd-Scuseria-Ernzerhof (HSE) hybrid functional [23]. In order to reproduce the experimental band gaps, we set the mixing parameter for the non-local Hartree-Fock exchange to 0.27 for $MgGa_2O_4$ and 0.34 for $MgAl_2O_4$. The cut-off energy for the plane wave basis was set at 500 eV and the atomic positions were fully relaxed until the residual forces were less than 0.01 eV/Å. All defect calculations were performed using a 112-atom supercell (8 eight primitive cells) and a single Γ point for the Brillouin zone integration. Lattice constants of $MgGa_2O_4$ and $MgAl_2O_4$ were optimized to be 8.457 Å, and 8.084 Å, respectively, which agree well with experimental results [24]. Based on our tests using larger supercells, the errors in transition levels are less than 0.1 eV.

The formation energy of a defect $\alpha$ in the charge state $q$ is given by

$$\Delta H(\varepsilon_f, \mu_\alpha) = E_q - E_H - \sum_\alpha n_\alpha (\mu_\alpha + \mu_\alpha^{ref}) + q(\varepsilon_{VBM} + \varepsilon_f), \qquad (1)$$

where $E_q$ is the total energy of the supercell containing defect in charge state $q$, and $E_H$ is the total energy of the defect-free supercell, $n_\alpha$ is the number of atoms added ($n > 0$) or removed ($n < 0$), $\mu_\alpha$ represents the chemical potential of atomic species $\alpha$, reference to $\mu_\alpha^{ref}$, which is the chemical potential of the atomic species in its elemental phase (bulk or gas)e. $\varepsilon_{VBM}$ is the energy of the valence band maximum (VBM) and $\varepsilon_f$ is the Fermi energy relative to the VBM.

The thermodynamic charge transition level of a defect, $\varepsilon(q/q')$, defined as the Fermi level at which the formation energies of two charge states $q$, and $q'$ are equal, is given by :

$$\varepsilon(q/q') = [\Delta H(q) - \Delta H(q')]/(q'-q). \qquad (2)$$

Optical transitions involving defect levels and the conduction/valence bands are calculated using the same methodology, but both $\Delta H(q)$ and $\Delta H(q')$ are evaluated using the relaxed defect structure in the initial state of the optical transition. For example, the excitation energy for promoting an electron from a defect in charge state $q$ to the conduction band minimum (CBM) is given by $E_g - \varepsilon_{opt}(q/q-1)$; here, $\varepsilon_{opt}(q/q-1)$ is calculated using the relaxed defect structure at the charge state $q$. For the subsequent radiative recombination, the emission energy is given by $E_g - \varepsilon_{opt}(q-1/q)$; here, $\varepsilon_{opt}(q-1/q)$ is calculated using the relaxed defect structure of the charge state $q-1$. For intradefect optical transitions, we use the $\Delta$ self-consistent field ($\Delta$SCF) method [27-29] by fixing the occupation numbers of the electron and hole-occupied eigenlevels. Charge Image and potential alignment corrections [25, 26] were performed wherever appropriate.

To evaluate radiative and nonradiative capture coefficients, we adopt the method developed by the Van de Walle group. This approach considers multiphonon emission by approximating the phonon vibrations through a single effective mode [30-32]. This method has been demonstrated to be accurate and efficient in determining nonradiative and radiative carrier capture coefficients at defects in semiconductors [30-33].

## III. Results and Discussion

### A. Electronic structures of MgGa$_2$O$_4$ and MgAl$_2$O$_4$

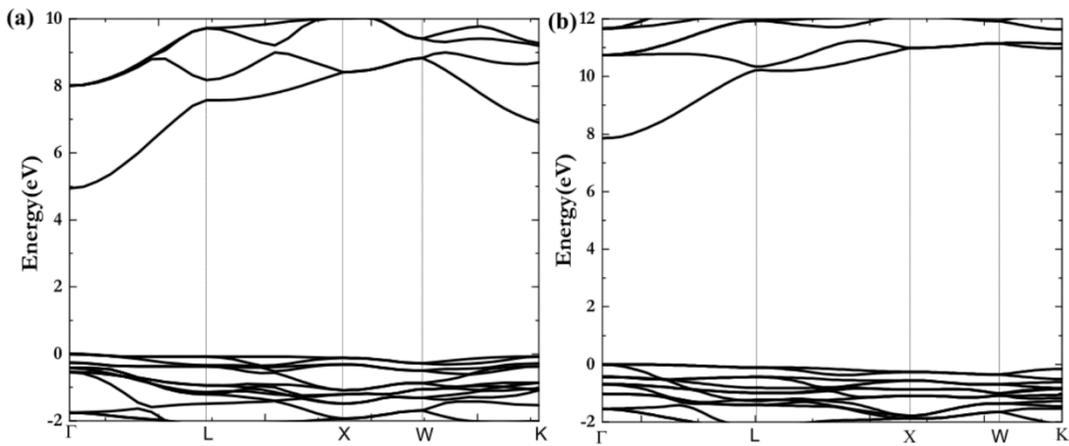

FIG. 1. Calculated band structures for (a) MgGa$_2$O$_4$ and (b) MgAl$_2$O$_4$ using the HSE hybrid functional.

Both MgGa$_2$O$_4$ and MgAl$_2$O$_4$ exhibit a direct band gap at the $\Gamma$ point. By tuning the fraction of exact exchange in the hybrid functional, we obtain band gap values of 4.94 for MgGa$_2$O$_4$ and and 7.75 eV for MgAl$_2$O$_4$. These results are in excellent agreement with the corresponding experimental values of 4.9 and 7.8 eV, respectively [34-36].

The calculated band structures of MgGa$_2$O$_4$ and MgAl$_2$O$_4$ are shown in Figure 1. For both materials, the valence band is mainly made up of O $2p$ states. In MgGa$_2$O$_4$, the conduction band is dispersive and dominated by delocalized Ga $s$ states, while in MgAl$_2$O$_4$, it consists of delocalized

Al *s* states. The valence bands in both cases are relatively flat, consistent with the localized nature of the O 2*p* orbitals.

## B.  Native point defects in MgGa$_2$O$_4$ and MgAl$_2$O$_4$

To calculate defect formation energies as defined in Eq. (1), the chemical potentials, which depend on growth conditions, must first be determined. Taking MgGa$_2$O$_4$ as an example, the chemical potentials of Mg, Ga, and O ($\mu_{Mg}$, $\mu_{Ga}$, $\mu_O$) must meet the following conditions to avoid the formation of competing phases,:

$$\mu_{Mg} + \mu_O \leq \Delta H_f(MgO)$$
$$2\mu_{Ga} + 3\mu_O \leq \Delta H_f(Ga_2O_3)$$
$$\mu_{Mg} + 2\mu_{Ga} + 4\mu_O = \Delta H_f(MgGa_2O_4) = -19.60 \text{eV},$$

where $\Delta H_f(MgGa_2O_4)$ is the enthalpy of formation of MgGa$_2$O$_4$. The chemical potentials are referenced to their elemental states (metallic Mg, Ga, and O$_2$ gas). Similar constraints apply for MgAl$_2$O$_4$. The calculated chemical potential phase diagrams are shown in Figure 2.

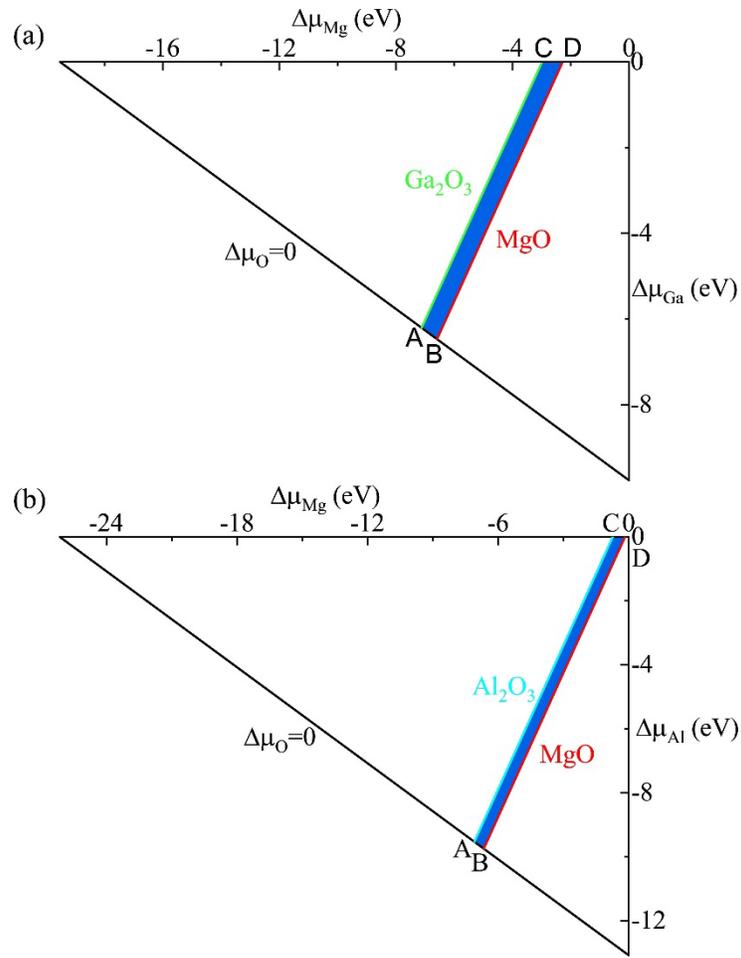

FIG. 2. Calculated phase diagrams for (a) MgGa$_2$O$_4$ and (b) MgAl$_2$O$_4$. The shaded polygon ABCD are the regions where MgGa$_2$O$_4$ and MgAl$_2$O$_4$ are stable without forming other phases.

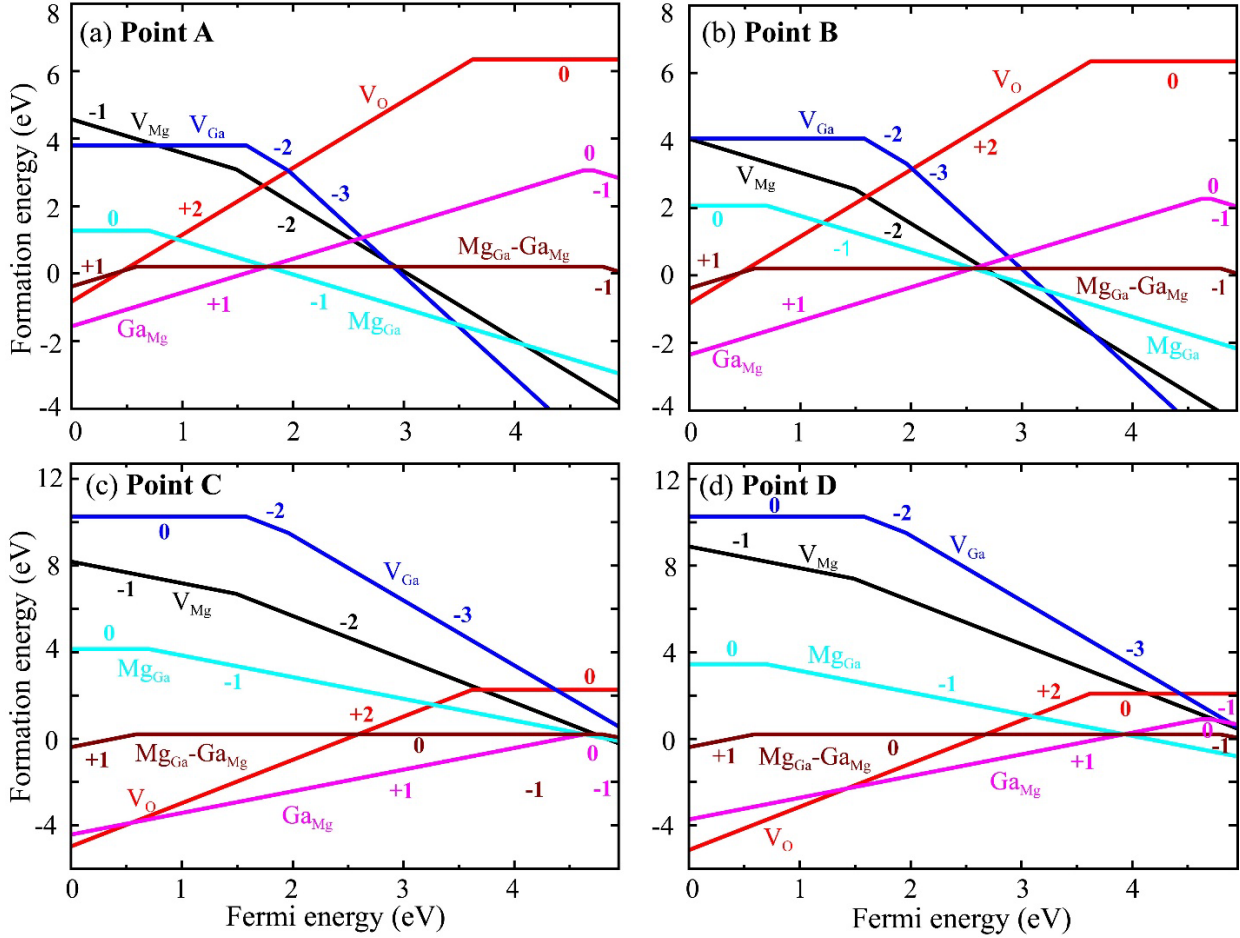

FIG. 3. Calculated formation energies of native defects in MgGa$_2$O$_4$ (a), (b), (c) and (d) using chemical potentials corresponding to points A, B, C, and D in Figure 2, respectively. Slopes of the formation energy lines indicate the charge states of the defects. A transition level is where the slope changes.

Figure 3 shows the formation energies of important native defects in MgGa$_2$O$_4$ calculated with chemical potentials corresponding to points A, B, C, and D in the phase diagram (Figure 2a). Only low-energy defects, such as antisite defects (Mg$_{Ga}$, Ga$_{Mg}$) and vacancies (V$_O$, V$_{Mg}$ and V$_{Ga}$), are included, While interstitial defects (Mg$_i$, Ga$_i$ and O$_i$) are omitted due to their high formation energies.

In the absence of a high concentration of impurities, the Fermi level should be close to the crossing point of the formation energy lines of the lowest-energy native donor and acceptor defects. At point A in Figure 3, the Fermi level is close to $\varepsilon_{VBM}$ + 1.77 eV, near the intersection of Mg$_{Ga}$ and Ga$_{Mg}$ lines, suggesting slightly *p*-type conductivity. At Point B, the Fermi energy lies at $\varepsilon_{VBM}$ + 2.56 eV, is close to midgap (2.47 eV), implying high resistivity. Both A and B correspond to oxygen-rich conditions. In contrast, Points C and D represent O-poor and Ga-rich conditions, with Fermi levels near $\varepsilon_{VBM}$ + 4.63 eV and $\varepsilon_{VBM}$ + 3.93 eV, respectively. These results show that MgGa$_2$O$_4$ exhibits a tunable transition between insulating and semiconducting behaviors depending on growth conditions. This result is consistent with experimental findings: melt-grown MgGa$_2$O$_4$ crystals have been reported as either insulators or semiconductors, depending on the oxygen content during growth [34].

In addition, inverse spinel configurations have been observed experimentally in MgGa$_2$O$_4$ [6], Implying that antisite complex Ga$_{Mg}$-Mg$_{Ga}$ should form readily. As expected, our calculated formation energy of Ga$_{Mg}$-Mg$_{Ga}$ complex is only 0.21 eV, with the complex being most stable in the neutral charge state, as showed in Figure 3.

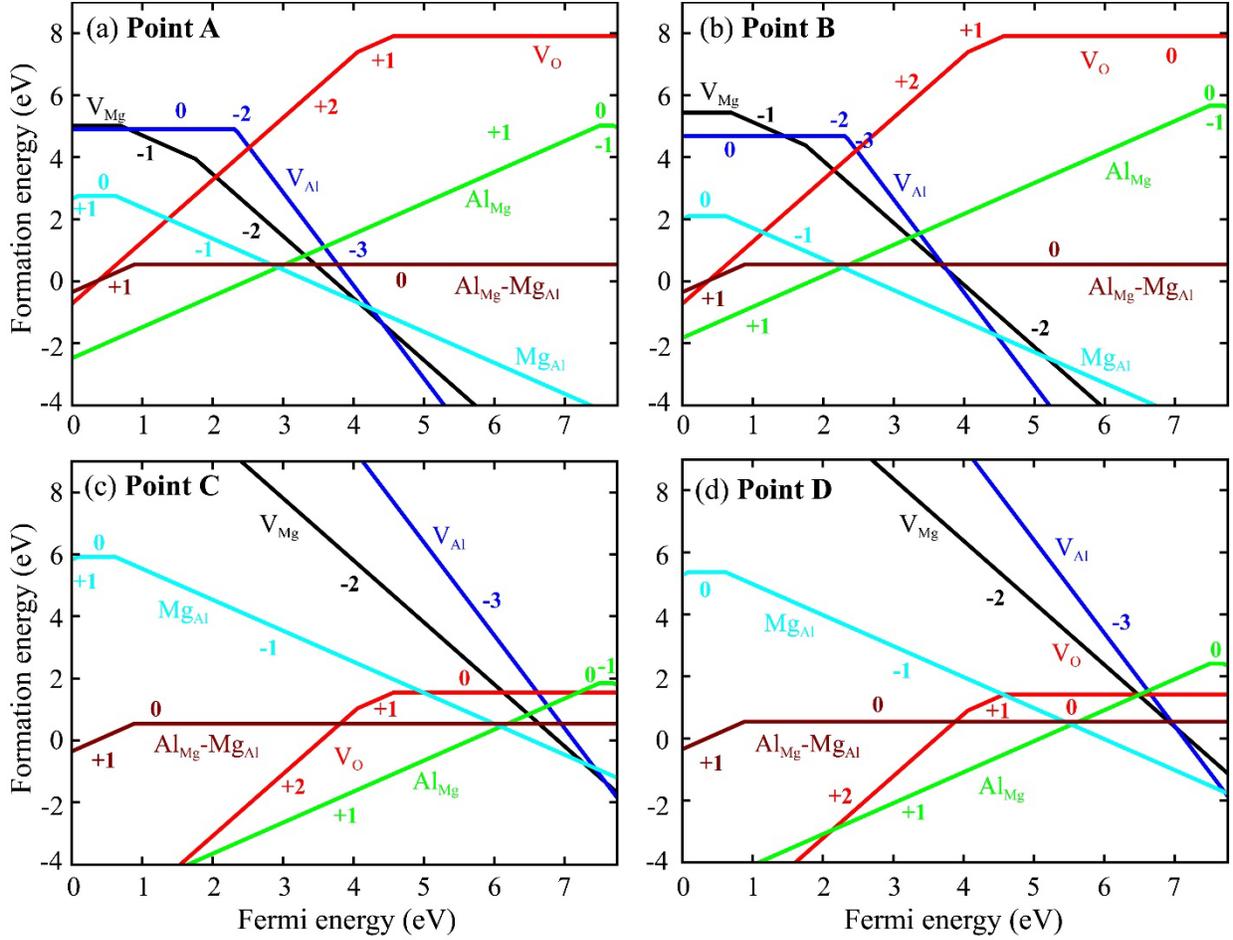

FIG. 4. Calculated formation energies of native defects in MgAl$_2$O$_4$. Slopes of the formation energy lines indicate the charge states of the defects. A transition level is where the slope changes.

Figures 4 presents the calculated formation energies for native point defects in MgAl$_2$O$_4$. Similar to MgGa$_2$O$_4$, vacancies (V$_O$, V$_{Mg}$, V$_{Al}$) and antisite defects (Mg$_{Al}$, Al$_{Mg}$) are the dominating defects. At the four chemical potential points (A-D) defined in Figure 2b, the Fermi levels lie near the crossing points of Mg$_{Al}$ and Al$_{Mg}$ formation energy lines: $\varepsilon_{VBM}$ + 2.92 eV (Point A), $\varepsilon_{VBM}$ + 2.27 eV (Point B), $\varepsilon_{VBM}$ + 6.09 eV (Point C), and $\varepsilon_{VBM}$ + 5.54 eV (Point D). Again, O-rich conditions (Points A and B) favor *p*-type behavior, while O-poor conditions (Points C and D) promote *n*-type conductivity, mirroring the behavior in MgGa$_2$O$_4$. The formation energy of the neutral antisite defect Al$_{Mg}$-Mg$_{Al}$ is also low, similar to the Ga$_{Mg}$-Mg$_{Ga}$ complex in MgGa$_2$O$_4$.

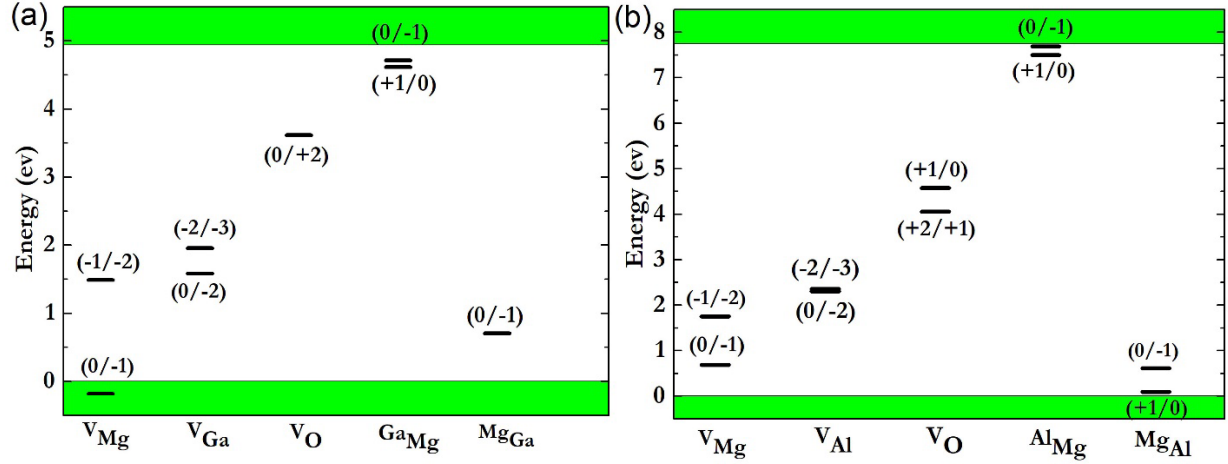

FIG. 5. Calculated thermodynamic charge transition levels for native defects in $MgGa_2O_4$ and $MgAl_2O_4$.

Based on the formation energies at different charge states, we calculate the thermodynamic charge transition levels, shown in Figure 5. For $MgGa_2O_4$, the transition level (0/2+) of $V_O$ is located at 3.62 eV above the VBM, indicating that $V_O$ is a very deep donor. $V_O$ is a negative U center with its +1 charge state metastable. In $MgGa_2O_4$, an O atom has one Mg and three Ga nearest neighbors, with Mg-O and Ga-O bond lengths of 2.01 Å and 2.02 Å, respectively. For $V_O$ in the 2+ state, the positively charged Mg and Ga ions repel each other, resulting in significant increased distance from the center of $V_O$ to Mg (2.77 Å) and Ga (2.26 Å), while for neutral $V_O$, the Mg and Ga ions surrounding the vacancy site move inward to the positions close to those in pristine $MgGa_2O_4$, and the two electrons are localized at the vacancy site. The antisite $Mg_{Ga}$ has a (-/0) transition level at 0.70 eV above the VBM, making it a deep acceptor, while $Ga_{Mg}$ acts as a donor with its (+/0) at 0.32 eV below the CBM. Cation vacancies $V_{Mg}$ and $V_{Ga}$ are deep acceptors with their (2-/-) and (3-/2-) hole trapping levels at 1.49 eV and 1.96 eV, respectively. The $Ga_{Mg}$-$Mg_{Ga}$ complex is amphoteric with its (0/+) and (0/-) transition levels located at $E_{VBM}$ + 0.59 eV and $E_{CBM}$ - 0.15 eV, respectively.

For MgAl$_2$O$_4$, as shown in Figure 5(b), Al$_{Mg}$ and V$_O$ are dominant donor defects. In contrast to V$_O$ in MgGa$_2$O$_4$, V$_O$ in MgAl$_2$O$_4$ is not a negative-U center, with (0/+) and (+/2+) deep donor levels at 3.18 and 3.69 eV below the CBM, respectively. An O atom in perfect MgAl$_2$O$_4$ is coordinated by three Al and one Mg atoms, with Al-O bonds of 1.92 Å and the Mg-O bond of 1.94 Å. In V$_O^0$, the Mg nearest neighbors relax inward by 0.15 Å, while the three Al-V$_O$ bonds remain unchanged. Upon ionization to +1 and +2 charge states, the Mg/Al nearest neighbors relax outward by 0.04/0.13 Å and 0.87/0.23 Å, respectively. Al$_{Mg}$ it is a deep donor with a (+/0) level at 0.25 eV below the CBM, and Mg$_{Al}$ is a deep acceptor with a (-/0) transition level at 0.62 eV above the VBM.

## C. Optical transitions in MgGa$_2$O$_4$ and MgAl$_2$O$_4$

As discussed before, since the long persistent luminescence properties of MgGa$_2$O$_4$ and MgAl$_2$O$_4$ are self-activated [6, 8], meaning they arise from intrinsic defects rather than extrinsic dopants. Therefore, identifying the defect-related optical transitions is key to understanding the luminescence mechanisms in these materials.

Building on our defect formation and charge transition calculations, we further investigate the optical transitions associated with intrinsic defects in MgGa$_2$O$_4$ and MgAl$_2$O$_4$. Optical transitions can take place between a defect level and the conduction or valence band. We have studied all possible optical transitions involving low-energy defects and report here the results that either align closely with experimental data or involve defects whose roles in persistent luminescence are debated.

### 1. Optical transitions in MgGa$_2$O$_4$

Figure 6 shows the configuration coordinate diagrams for optical transitions associated with $V_O$ and $Mg_{Ga}$ in $MgGa_2O_4$.

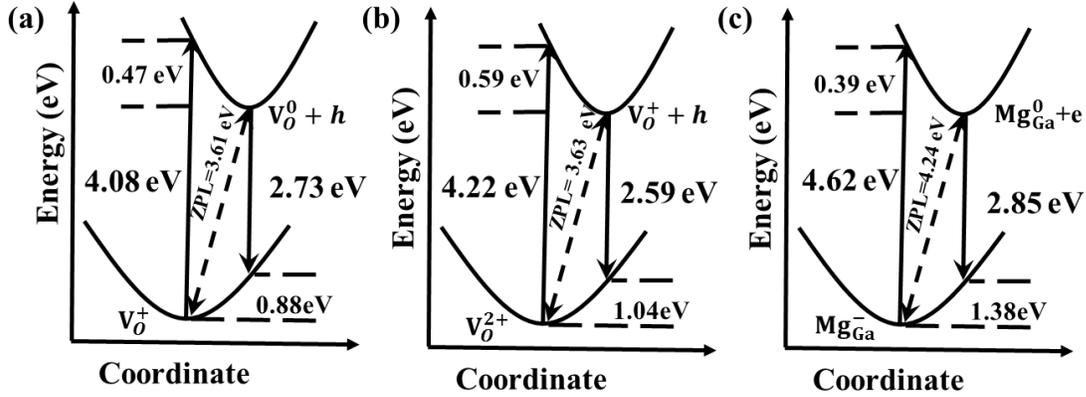

FIG. 6. Schematic configuration coordinate diagrams for optical transitions associated with (a) the $V_O$ (+/0) level exchanging a hole with the VBM, (b) the $V_O$ (+/2+) level exchanging a hole with the VBM, (c) the $Mg_{Ga}$ (-/0) level exchanging an electron with CBM.

Figure 3 shows that $V_O$ in $MgGa_2O_4$ is a negative-U center: only the neutral or 2+ charged state can be stable depending on the O chemical potential, while the +1 charge state is metastable and can be populated via optical excitation. Under the O-poor condition, $V_O$ is neutral. Under optical excitation, free holes can be generated in the valence band and recombine with electrons localized at neutral oxygen vacancies. Figure 6(a) shows that the emission energy associated with this recombination process ($V_O^0 + h \rightarrow V_O^+$) is 2.73 eV, with a zero-phonon line at 3.61 eV. The corresponding vertical excitation energy for the reverse process (hole excitation from $V_O^+$ to the valence band) is 4.08 eV. These results are in good agreement with experimental observations: (1) An emission peak at 2.76 eV under 235 nm (5.28 eV) band-to-band excitation is attributed to oxygen vacancies [15]; (2) A diffuse reflectance absorption band at 300 nm (4.13 eV) agrees well with our calculated absorption peak at 4.08 eV, supporting the assignment to $V_O$ absorption [15].

The optical transition associated with the $V_O$ (+/2+) level is shown in Figure 6(b). An emission energy of 2.59 eV and a ZPL of 3.63 eV associated with the transition $V_O^+ + h \rightarrow V_O^{2+}$ are obtained. However, this process is expected to be inefficient because it involves hole capture by a positively charged defect.

Another important intrinsic defect is $Mg_{Ga}$, whose optical transition diagram is shown in Figure 6(c). For optical excitation from $Mg_{Ga}^-$ to the CBM ($Mg_{Ga}^- \rightarrow Mg_{Ga}^0 + e$), we obtain an optical excitation energy of 4.62 eV, which agrees well with the experimental 265 nm excitation used in the photoluminescence excitation (PLE) measurements [6]. The corresponding recombination process ($Mg_{Ga}^0 + e \rightarrow Mg_{Ga}^-$) involves the recombination of electrons from the conduction band with holes in the defect state. Our calculated emission energy is 2.85 eV with a large relaxation energy 1.38 eV, which is consistent with the broad band emission peak at 424 nm (2.92 eV) observed in experiment [6]. The calculated ZPL of 4.24 eV is also in good agreement with the PLE onset at about 280 nm (4.43 eV) [6].

As shown in Figure 3, the formation energy of $Mg_{Ga}^-$ is very low, particularly under the oxygen rich condition. Therefore, the transitions related to $Mg_{Ga}^-$ could be the source of the stronger PL peak at about 2.9 eV [6]. Similarly, the formation energy of neutral oxygen vacancy is also relative low under the oxygen poor condition. Therefore, both $V_O$ and $Mg_{Ga}$ may contribute to the broad blue emission band observed in MgGa$_2$O$_4$.

We also investigated donor-acceptor complexes, specifically $V_O + V_{Mg}$ and $V_O + V_{Ga}$, which have been proposed as possible origins of the observed PL peaks [3, 6]. For the $V_O + V_{Mg}$ complex, only the neutral charge state is considered. The calculated emission energy is 1.20 eV. For the $V_O + V_{Ga}$ complex, two charge states states (neutral and -1) are examined. The corresponding emission energies are calculated to be 1.67 and 1.14 eV, respectively. Therefore, $V_O + V_{Mg}$ and $V_O + V_{Ga}$ complexes are not responsible for the experimental PL peaks above 2 eV [3, 6, 14]. However, (V$_O$ + V$_{Ga}$)$^-$ could potentially contribute to the red emission around 708 nm (1.75 eV) observed in experiment [13].

## 2. Optical transitions in MgAl$_2$O$_4$

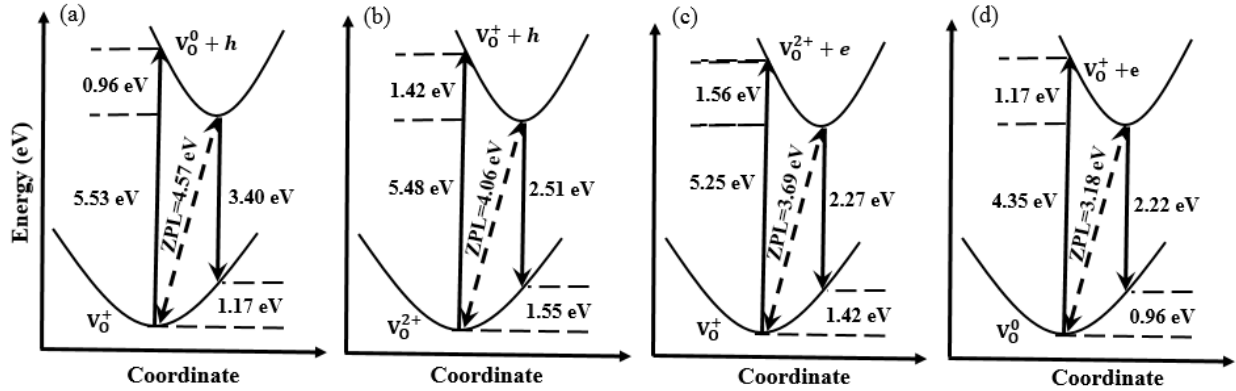

FIG. 7. Schematic configuration coordinate diagrams for optical transitions associated with the (a) V$_O$ (+/0) level exchanging a hole with the VBM, (b) V$_O$ (+/2+) level exchanging a hole with the VBM, (c) V$_O$ (+/2+) level exchanging an electron with the CBM, (d) V$_O$ (+/0) level exchanging an electron with the CBM.

Figure 7 shows schematic configuration coordinate diagrams of various optical transitions involving V$_O$ in MgAl$_2$O$_4$. Unlike V$_O$ in MgGa$_2$O$_4$, V$_O$ in MgAl$_2$O$_4$ can exist in three charged states: 0, +1, and +2. This enables a greater variety of optical transitions. For instance, if the ground state of V$_O$ is neutral, as suggested by our formation energy results (Figure 4(c) and 4(d)), then below-bandgap excitation can promote trapped electrons to the conduction band. Also, neutral V$_O$ can capture holes generated via cross-gap excitation.

Figure 7(a) depicts the excitation of hole from $V_O^+$ to the valence band, resulting in $V_O^0$ and a free hole, and the reverse recombination process. The emission energy due to the recombination between $V_O^0$ and a free hole is calculated to be 3.40 eV and the ZPL is at 4.57 eV.

Figure 7(b) shows a similar hole excitation from $V_O^{2+}$ to $V_O^+$. Our calculated absorption and emission energies are 5.48 and 2.51 eV, respectively. This process is expected to be less efficient, as it involves hole capture by a positively charged defect. Figure 7 (c) shows the elelction excitation from $V_O^+$ to CBM and the subsequent recombination between $V_O^{2+}$ and a free electron. Our calculated absorption peak is at 5.25 eV, while the emission peak is at 2.27 eV with a ZPL of 3.69 eV. Figure 7 (d) shows a similar elelction excitation from $V_O^0$ to the CBM and the recombination between $V_O^+$ and a free electron. Our calculated excitation and emission energies are 4.35 eV and 2.22 eV, respectively.

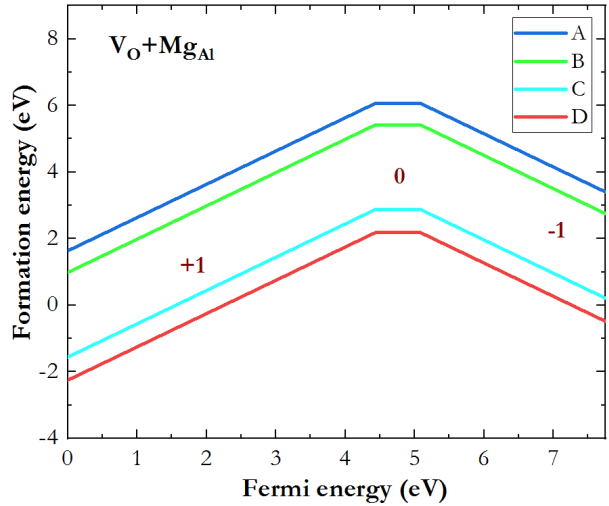

FIG. 8. Calculated formation energies of $V_O$-$Mg_{Al}$ complex in MgAl$_2$O$_4$ using chemical potentials corresponding to points A, B, C, and D in Figure 2, respectively.

We also investigated optical transitions in donor-acceptor complex $V_O$-$Mg_{Al}$, where a dominant donor V$_O$ binds with an acceptor Mg$_{Al}$. The complex can exist in positively charged ($V_O$-$Mg_{Al}$)$^+$, neutral ($V_O$-$Mg_{Al}$)$^0$ or negatively charged ($V_O$-$Mg_{Al}$)$^-$, as the Fermi level varies within the band gap. Taking ($V_O$-$Mg_{Al}$)$^0$ as an example, our calculated binding energy between $V_O^+$ and $Mg_{Al}^-$ is -4.59 eV, indicating a highly stable configuration due to charge transfer, Coulomb attraction, and atomic relaxation. Figure 8 shows the formation energies of these complexes across the chemical potential points from Figure 2.

Figure 9 illustrates the configuration coordinate diagrams for intradefect transitions of the $V_O$-$Mg_{Al}$ complex. For the neutral state $(V_O$-$Mg_{Al})^0$, the excitation and emission energies are calculated to be 5.42 eV and 2.60 eV, respectively (Figure 9a). The single-particle energy levels for the ground and excited states of $(V_O$-$Mg_{Al})^0$ are shown in Figure 10. For the positively charged complex $(V_O$-$Mg_{Al})^+$, the excitation and emission energies are 5.72 eV and 1.94 eV, respectively (Figure 9b).

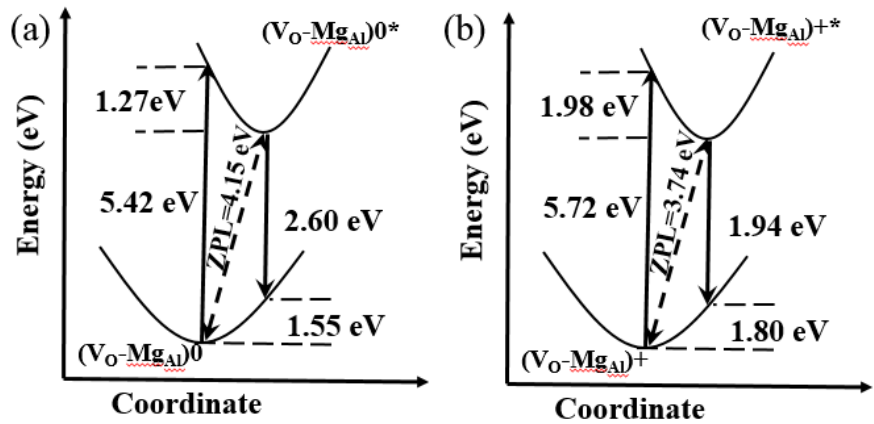

FIG. 9. Schematic configuration coordinate diagrams for the intradefect optical transitions associated with the donor-acceptor complex $V_O$-$Mg_{Ga}$ (a) in neutral state, (b) in positive state.

In addition, electron capture from the CBM by $V_O^{2+}$-$Mg_{Al}^-$ results in $V_O^+$-$Mg_{Al}^-$, with an emission energy of 1.84 eV, which is lower than the 2.27 eV emission due to the electron capture from the CBM by an isolated $V_O^{2+}$ (Figure 7c). This shift reflects the shallower defect level caused by the defect-defect interaction within the complex as showed in Figure 11. Our transition level analysis further support this result. The (0/-) level (the electron capture level) of the complex is at 0.07 eV below the CBM, shallower than the 0.37 eV (+/0) level of the isolated $V_O$. The (2+/+) level (the hole capture level) of the complex is 0.52 eV above the VBM, compared to 0.62 eV (0/-) level for isolated $Mg_{Al}$. The emission process of hole (at VBM) captured by negative state $(V_O\text{-}Mg_{Al})^-$ is not showed here due to the shallower level of $Mg_{Al}$ within $V_O$-$Mg_{Al}$ complex.

Additional donor-acceptor complexes such as $V_O$-$V_{Mg}$ and $V_O$-$V_{Al}$ were also investigated. However, their emission energies, 1.08 eV and 1.75 eV, suggest that they are not responsible for the experimentally observed PL peak at 2.7 eV [4, 7, 21].

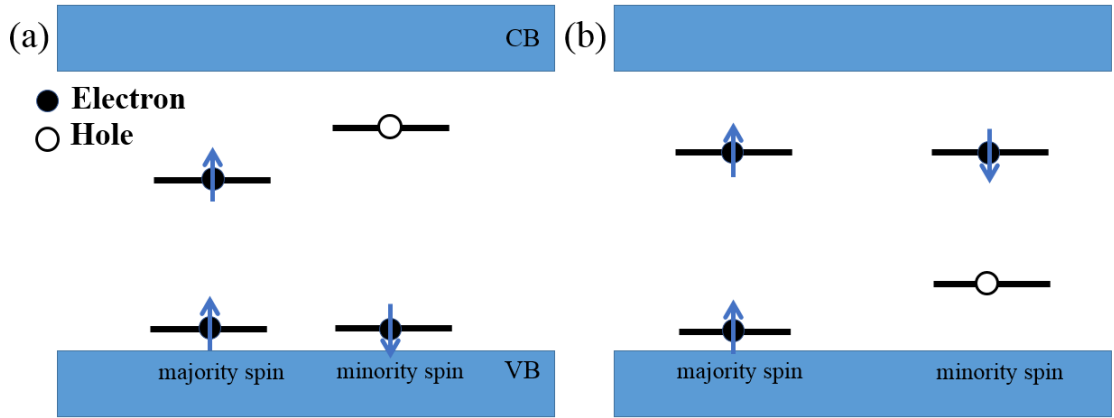

FIG. 10. Single-particle energy levels of $V_O$-$Mg_{Al}$ donor-acceptor complex for neutral states in $MgAl_2O_4$ (a) ground state and (b) excited state.

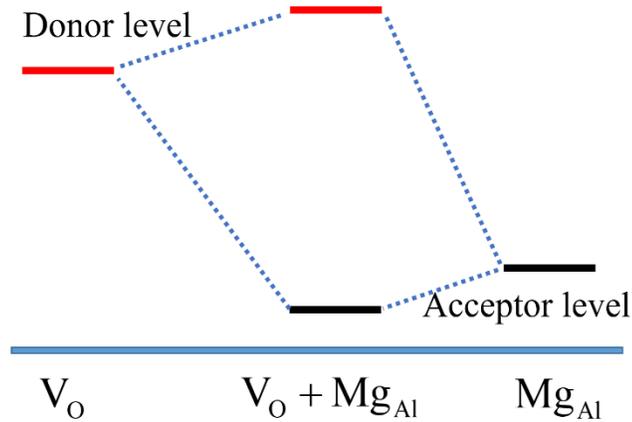

FIG. 11. Schematic of the defect-defect interaction $V_O$ and $Mg_{Al}$ within $V_O$-$Mg_{Al}$ complex.

The interaction makes both donor and acceptor levels shallower.

With the calculated optical properties of defects presented above, we now discuss the origin of the experimentally observed absorption and emission in $MgAl_2O_4$. Experimentally, oxygen vacancies are widely regarded as the origin of PL peaks at 2.7 or 2.35 eV [4, 7, 21]. For the emission peak of 2.7 eV, Sawai attributed it to $V_O^+$ [7], while Bandyopadhyay suggested a neutral $V_O$ [4]. For the peak of 2.35 eV, Raj proposed oxygen-related vacancies as the source [21]. As for excitation, two peaks centered at about 4.8 and 5.3 eV have been ascribed to $V_O^+$ and $V_O$, respectively [7]. Bandyopadhyay observed two peaks at 4.45 eV (unidentified electron trap) and 5.3 eV ($V_O$ center absorption) [4].

Due to the varied synthesis conditions, the charge state of the ground state oxygen vacancy can be +2, +1 or 0 [4, 7, 21], consistent with our defect calculations. As shown in Figure 7, transitions from $V_O^0$ and $V_O^+$ to the CBM require 4.35 and 5.25 eV, respectively, while the corresponding emission energies are 2.22 and 2.27 eV, both lower than the experimental emission peak at 2.7 eV [4, 7, 21].

Our calculations show that $V_O^0$ has a higher-lying occupied defect level than $V_O^+$, due to electron-electron repulsion. This results in the excitation energy of $V_O^0$ is lower than that for $V_O^+$, a trend opposite to that reported by Bandyopadhyay [4]. Based on our results, we propose the following assignments: (1) The absorption peak near 4.5 eV originates from $V_O^0$; (2) The absorption at 5.3-5.4 eV originates from $V_O^+$ and ($V_O^+$-$Mg_{Al}^-$); (3) The green emission stems from $V_O^+$ and $V_O^{2+}$; (4) The 2.7 eV emission is attributed to $V_O^+$-$Mg_{Al}^-$. These assignments are further supported by our calculated radiative capture coefficients, discussed below, which confirm that these defects serve as luminescent centers.

## 3. Radiative capture coefficients at deep defects in MgGa$_2$O$_4$ and MgAl$_2$O$_4$ at room temperature

Carrier recombination through defect centers in semiconductors can occur in both radiative and nonradiative pathways [30-33, 37, 38]. Typical radiative capture coefficients are on the order of $\sim 10^{-14} - 10^{-13}$ cm$^3$s$^{-1}$, while nonradiative capture coefficients can span a much wider range of $\sim 10^{-14} - 10^{-6}$ cm$^3$s$^{-1}$ [30].

To further characterize optical properties of Mg$_{Ga}$ and $V_O + Mg_{Al}$ as potential luminescent centers, we calculated their radiative capture coefficients using the methodology in Ref. [31]. The expression used is

$$C_n = f\eta_{sp} V \frac{e^2 n_r}{3m^2 \varepsilon_0 \pi c^3 \hbar^2} |p_{ij}|^2 E_{opt} = (5.77 \times 10^{-17})(V f n_r \eta_{sp} E_{opt} \frac{|p_{ij}|^2}{2m}) \text{cm}^3\text{s}^{-1},$$

where V is the volume of the supercell, $f$ is the Sommerfeld factor (1 for neutral centers in both cases), $n_r$ is the refractive index, $\eta_{sp}$ is the factor of spin selection rule (0.5 for the transitions such as doublet to singlet and triplet to doublet), $E_{opt}$ is the vertical optical transition energy, $|p_{ij}|^2$ is the square of momentum matrix element between the initial and final states involved in the optical transition. Our calculated $|p_{ij}|^2$ are 0.015 a.u. for Mg$_{Ga}$ and 0.029 a.u. for $V_O + Mg_{Al}$, leading to radiative electron capture coefficients of 1.3×10$^{-13}$ for Mg$_{Ga}$ and 0.5×10$^{-13}$ cm$^3$s$^{-1}$ for $V_O + Mg_{Al}$ at room temperature. These values fall within the typical range for radiative capture coefficients on the order of order ~10$^{-14}$-10$^{-13}$ cm$^3$s$^{-1}$ in semiconductors.

We also calculated the nonradiative capture coefficients using the multiphonon emission model described in Ref. [30]. For both Mg$_{Ga}$ and V$_O$ + Mg$_{Al}$, the nonradiative coefficients are much smaller than their radiative counterparts, confirming that these centers are highly efficient luminescent centers.

Furthermore, we evaluated the radiative and nonradiative carrier capture coefficients for V$_O$ in both MgGa$_2$O$_4$ and MgAl$_2$O$_4$. For MgGa$_2$O$_4$, the radiative capture coefficient for hole captured by the neutral V$_O$ (Figure 6(a)) is $3.2\times10^{-16}$ cm$^3$s$^{-1}$, while the corresponding nonradiative capture coefficient is insignificant. For MgAl$_2$O$_4$, the calculated radiative and nonradiative capture coefficients for electron capture by V$_O^+$ are $4.0\times10^{-13}$ and $6.1\times10^{-15}$ cm$^3$s$^{-1}$, respectively. Again, the radiative pathway is dominant. In contrast, for electron captured by V$_O^{2+}$, the nonradiative capture coefficient is $1.8\times10^{-4}$ cm$^3$s$^{-1}$, much higher than the radiative capture coefficient of $2.7\times10^{-13}$ cm$^3$s$^{-1}$ due to the small "classical" barrier [30]. For hole capture by neutral V$_O$, the radiative capture coefficient is $2.6\times10^{-13}$ cm$^3$s$^{-1}$, while the nonradiative contribution is negligible.

## 4. Carrier trap centers in MgGa$_2$O$_4$ and MgAl$_2$O$_4$

Trap centers can capture electrons from the conduction and holes from the valance band, temporarily storing excitation energy. Once external excitation stops, the captured electrons or holes can be released, primarily by thermal energy, and subsequently migrate to luminescent centers, where delayed recombination produces persistent luminescence [9]. The trap depth (*i.e.*, energy separation from the band edge) plays critical role in determining the duration of persistent luminescence at a given temperature. It has been reported that trap depths between 0.4 eV and 0.7 eV are essential for enabling persistent luminescence at room temperature [6]. Due to the presence of various defects, including vacancies, cation-cation antisites, and defect complexes, multiple trap centers with different depths may exist in MgGa$_2$O$_4$ and MgGa$_2$O$_4$.

As discussed in Section C 1 and 3, $Mg_{Ga}$ is a likely emission center for the observed blue emission in MgGa$_2$O$_4$. Based on our defect calculations, $V_O^{2+}$ and $Ga_{Mg}^{+}$ are stable under the intermediate oxygen chemical potential conditions (Figure 3). The donor defect Ga$_{Mg}$ has a (+/0) transition level at 0.32 eV below the CBM, making it an electron trap that can release electrons slowly at room temperature. In contrast, the trap level of V$_O$ lies at 1.32 eV below the CBM, which is much deeper than that of Ga$_{Mg}$ and thus releases electrons more slowly, contributing to longer-duration luminescence. Therefore, in MgGa$_2$O$_4$, Mg$_{Ga}$ and Ga$_{Mg}$ (or V$_O$) may act as emission centers and carrier traps, respectively, supporting persistent luminescence.

In addition, donor-acceptor pairs involving V$_O$, such as V$_O$+V$_{Mg}$ and V$_O$+V$_{Ga}$, can also act as effective trap centers for persistent luminescence in MgGa$_2$O$_4$. Due to the repulsion interactions between defect levels within the defect complex (Figure 11), the defect level of V$_O$ is pushed closer to the CBM within these complexes, making carrier release more thermally accessible. For V$_O$+V$_{Mg}$, our calculation show that the (0/2-) transition level is at 0.73 eV below the CBM and the (0/-) level is at 0.65 eV below the CBM. For V$_O$+V$_{Ga}$, the (-/3-) and (-/2-)levels are pushed higher (relative to the corresponding levels of isolated V$_O$) to be 0.48 eV and 0.39 eV below the CBM, respectively. Consequently, the electrons trapped by the V$_O$ levels in V$_O$+V$_{Mg}$ and V$_O$+V$_{Ga}$ defect complexes can be emitted more easily than those trapped by isolated V$_O$ defects. These results agree well with thermoluminescence spectroscopy measurements, which report two electron traps with approximate depths of 0.65 eV and 1.23 eV [15]. That study also suggests a multi-step detrapping pathway, where electrons are first promoted from deep traps to shallow ones, and then to the conduction band, a mechanism that supports our theoretical findings.

For MgAl$_2$O$_4$, an isolated Mg$_{Al}$ is a deep acceptor with a (0/-) transition level located at 0.62 eV above the VBM. Holes can be thermally released from this trap and recombine at nearly luminescence centers, giving rise to persistent luminescence. Meanwhile, $Al_{Mg}$ is a donor with a (0/+) transition level at 0.25 eV below the CBM, which can act an electron trap. Figure 12 provide schematic illustrations of the trapping-detrapping mechanism responsible for persistent luminescence in $MgGa_2O_4$ and $MgAl_2O_4$. For clarity, only isolated point defect trap centers are showed.

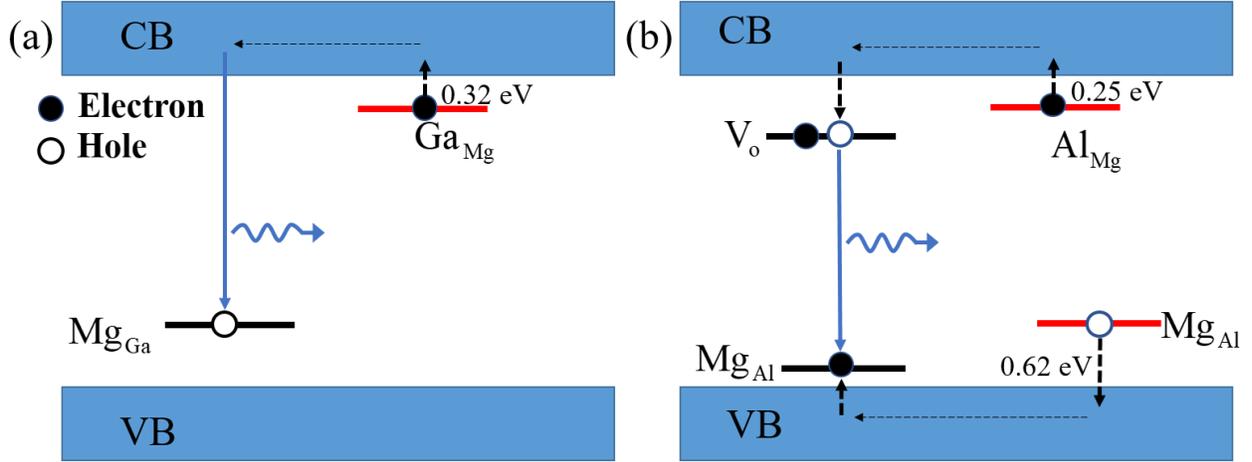

FIG. 12. Schematic illustrations of (a) electron trapping-detrapping model in $MgGa_2O_4$, and (b) the hole and electron trapping-detrapping model in $MgAl_2O_4$. Only isolated point defect trap centers are shown.

## IV. Conclusion

We have conducted an in-depth theoretical investigation of the self-activated persistent luminesce properties of $MgGa_2O_4$ and $MgAl_2O_4$, focusing on roles of intrinsic defects as luminescent centers and carrier traps. Using hybrid functional calculations, we systematically analyzed the formation energies, thermodynamic charge transition levels, and optical transition processes associated with native defects in both materials. Our results reveal that cation antisite defects and oxygen vacancies are the most important native defects in both materials. These defects introduce deep levels within the band gap, enabling them to act as both luminescent centers and carrier traps.

Our calculated optical transition energies suggest that ansite $Mg_{Ga}$ is a strong candidate for the observed blue emission in $MgGa_2O_4$, and the $V_O$+$Mg_{Al}$ defect complex is likely responsible for the dominant emission peak at 2.7 eV in $MgAl_2O_4$. Our calculated radiative electron capture coefficients for these defects fall within the expected range for semiconductors, further supporting their roles as efficient luminescent centers.

We also examined the role of oxygen vacancy as both absorption and emission centers in $MgAl_2O_4$, Moreover, we demonstrated that isolated $V_O$ as well as donor-acceptor complexes ($V_O$-$V_{Mg}$ and $V_O$-$V_{Ga}$) can serve as effective carrier trap centers, contributing to the delayed emission, which is critical for persistent luminescence.

Overall, from the perspective of intrinsic defect physics, our comprehensive and detailed study offers valuable insights into the microscopic origin of self-activated persistent luminescence in spinel oxides. These findings provide a theoretical foundation for the design and optimization of new, efficient dopant-free phosphors for novel applications.

**ACKNOWLEDGMENTS**
This work was supported by the National Natural Science Foundation of China (NSFC) under Grants No. 12174017, No. 12374079, and No.11604007. The work at Jinan University was supported by the Natural Science Foundation of Shandong Province (ZR2024QA137) and Annual New Talents Research Program of University of Jinan-Excellent Doctoral Project (XBS2475).

# References:

[1] T. Maldiney, A. Bessière, J. Seguin, E. Teston, S. Sharma, B. Viana, A. Bos, P. Dorenbos, M. Bessodes, D. Gourier, D. Scherman, C. Richard, The in vivo activation of persistent nanophosphors for optical imaging of vascularization, tumours and grafted cells, Nature Materials. **13**, 418 (2014).

[2] J. Botterman and P. F. Smet, Persistent phosphor SrAl$_2$O$_4$:Eu,Dy in outdoor conditions: saved by the trap distribution, Opt. Express. **23**, A868 (2015).

[3] Z. Liu, P. Hu, X. Jing, and L. Wang, Luminescence of Native Defects in MgGa$_2$O$_4$, J. Electrochem. Soc. **156**, H43 (2009).

[4] P. K. Bandyopadhyay and G. P. Summers, Luminescence and photoconductivity in magnesium aluminum spinel, Phys. Rev. B. **31**, 2422 (1985).

[5] G. J. Pogatshnik and Y. Chen and L. S. Cain, Optical transitions in neutron-irradiated MgAl$_2$O$_4$ spinel crystals, Phys. Rev. B. **37**, 2645 (1988).

[6] B. Jiang, F. Chi, X. Wei, Y. Chen, and M. Yin, A self-activated MgGa$_2$O$_4$ for persistent luminescence phosphor, J. Appl. Phys. **124**, 063101 (2018).

[7] S. Sawai and T. Uchino, Visible photoluminescence from MgAl$_2$O$_4$ spinel with cation disorder and oxygen vacancy, J. Appl. Phys. **112**, 258 (2012).

[8] N. Pathak, B. Sanyal, S. K. Gupta, and R.Kadam, MgAl$_2$O$_4$ both as short and long persistent phosphor material: role of antisite defect centers in determining the decay kinetics, Solid state Sciences **88**, 13 (2019).

[9] J. Xu and S. Tanabe, Persistent luminescence instead of phosphorescence: History, mechanism, and perspective, J. Lumin. **205**, 581 (2019).

[10] J. Liu, X. Duan, Y. Zhang, and H. Jiang, Cation distribution and photoluminescence properties of Mn-doped ZnGa$_2$O$_4$ nanoparticles, J. Phys. Chem. Solids. **81**, 15 (2015).

[11] M. Allix, S. Chenu, E. Véron, T. Poumeyrol, E. A. Kouadri-Boudjelthia, S. Alahraché, F. Porcher, D. Massiot, and F. Fayon, Considerable Improvement of Long-Persistent Luminescence in Germanium and Tin Substituted ZnGa$_2$O$_4$, Chem. Mater. **25**, 1600 (2013).

[12] S. N. Basavaraju, Storage of Visible Light for Long-Lasting Phosphorescence in Chromium-Doped Zinc Gallate, Chem. Mater. **26**, 1365 (2014).

[13] T. Yang, C. Shou, J. Tran, A. Almujtabi, Q. S. Mahmud, E. Zhu, Y. Li, P. Wei, and J. Liu, Photoluminescence study of MgGa$_2$O$_4$, Appl. Phys. Lett. **125**, 071903 (2024).

[14] G. Wang, T. Wang, Y. Yue, L. Hou, W. Huang, X. Zhu, H. Liu, and X. Yu, Regulation of Intrinsic Defects of Self-Activated MgGa$_2$O$_4$ Phosphors for Temperature Dynamic Anti-Counterfeiting, Laser. Photonics. Rev. **18**, 2400793 (2024).

[15] D. V. Mlotswa, L. L. Noto, S. J. Mofokeng, K. O. Obodo, V. R. Orante-Barrón, and B. M. Mothudi, Luminescence dynamics of MgGa$_2$O$_4$ prepared by solution combustion synthesis, Opt. Mater. **109**, 110134 (2020).

[16] S. Jiang, T. Lu, J. Zhang, and J. Chen, First-principles study on the effects of point vacancies on the spectral properties of MgAl$_2$O$_4$, Solid. State. Commun. **151**, 29 (2011).

[17] S. Jiang, T. Lu, Y. Long, and J. Chen, Ab initio many-body study of the electronic and optical properties of MgAl$_2$O$_4$ spinel, J. Appl. Phys. **111**, 43516 (2012).

[18] N. Pathak, P. S. Ghosh, S. K. Gupta, S. Mukherjee, R. M. Kadam, and A. Arya, An Insight into the Various Defects Induced Emission in MgAl$_2$O$_4$ and Their Tunability with Phase Behavior: Combined Experimental and Theoretical Approach, J. Phys. Chem. C. **120**, 4016 (2016).

[19] L. Museur, E. Feldbach, A. Kotlov, M. Kitaura, and A. Kanaev, Donor-acceptor pair transitions in MgAl$_2$O$_4$ spinel, J. Lumin. **265**, 8 (2024).

[20] G. P. Summers, G. S. White, K. H. Lee, and J. H. J. Crawford, Radiation damage in MgAl$_2$O$_4$, Phys. Rev. B. **21**, 2578 (1980).

[21] S. S. Raj, S. K. Gupta, V. Grover, K. P. Muthe, V. Natarajan, and A. K. Tyagi, MgAl$_2$O$_4$ spinel: Synthesis, carbon incorporation and defect-induced luminescence, J. Mol. Struct. **1089**, 81 (2015).

[22] J. Furthmüller and G. Kresse, Efficient iterative schemes for ab initio total-energy calculations using a plane-wave basis set, Phys. Rev. B. **54**, 11169 (1996).

[23] J. Heyd and G. E. Scuseria and M. Ernzerhof, Hybrid functionals based on a screened Coulomb potential, J. Chem. Phys. **118**, 8207 (2006).

[24] P. G. Casado, and I. Rasines, Crystal data for the spinels MGa$_2$O$_4$ (M = Mg, Mn), Zeitschrift für Kristallographie, **160**, 33 (1982).

[25] F. Christoph, J. Neugebauer and C. G. V. D. Walle, Fully ab initio finite-size corrections for charged-defect supercell calculations, Phys. Rev.Lett. **102**, 16402 (2009).

[26] C. Freysoldt and J. Neugebauer and C. G. V. D. Walle, Electrostatic interactions between charged defects in supercells, Phy. Status. Solidi. **248**, 10 (2011).

[27] R. O. Jones and O. Gunnarsson, The density functional formalism, its applications and prospects, Rev. Mod. Phys. **61**, 689 (1989).

[28] G. Rling and Andreas, Density-functional theory beyond the Hohenberg-Kohn theorem, Phys. Rev. A. **59**, 3359

[29] A. Hellman and B. Razaznejad and B. I. Lundqvist, Potential-energy surfaces for excited states in extended systems, J. Chem. Phys. **120**, 4593 (2004).

[30] A. Alkauskas, Q. Yan and D. W. C. G. Van, First-principles theory of nonradiative carrier capture via multiphonon emission, Phys. Rev. B. **90**, 075202 (2014).

[31] A. Alkauskas, C. E. Dreyer, J. L. Lyons, and D. W. C. G. Van, Role of excited states in Shockley-Read-Hall recombination in wide-band-gap semiconductors, Phys. Rev. B. **93**, 201304 (2016).

[32] D. Wickramaratne, C. E. Dreyer, J. X. Shen, J. L. Lyons, A. Alkauskas, and C. G. V. D. Walle, Deep-Level Defects and Impurities in InGaN Alloys, Phys. Status solidi B 257, 1900534 (2020).

[33] C. E. Dreyer, A. Alkauskas, J. L. Lyons, and C. G. Van de Walle, Radiative capture rates at deep defects from electronic structure calculations, Phys. Rev. B. **102**, 085305 (2020).

[34] Z. Galazka, D. Klimm, K. Irmscher, R. Uecker, M. Pietsch, R. Bertram, M. Naumann, M. Albrecht, A. Kwasniewski, R. Schewski, $MgGa_2O_4$ as a new wide bandgap transparent semiconducting oxide: growth and properties of bulk single crystals, Phys. Status Solidi A **212**, 1455 (2015).

[35] M. L. Bortz, R. H. French, D. J. Jones, R. V. Kasowski, and F. S. Ohuchi, Temperature dependence of the electronic structure of oxides: MgO, $MgAl_2O_4$ and $Al_2O_3$, Phys. Scripta. **41**, 537 (2006).

[36] B. Thielert, C. Janowitz, Z. Galazka, and M. Mulazzi, Theoretical and experimental investigation of the electronic properties of the wide band-gap transparent semiconductor $MgGa_2O_4$, Phys. Rev. B. **97** (2018).

[37] L. Shi, K. Xu and L. W. Wang, Comparative study of ab initio nonradiative recombination rate calculations under different formalisms, Phys. Rev. B. **91**, 51 (2015).

[38] H. Zhang, L. Shi, X. Yang, Y. Zhao, K. Xu, and L. Wang, First-principles calculations of quantum efficiency for point defects in semiconductors: the Example of Yellow Luminance by GaN: $C_N+O_N$ and GaN:$C_N$, Adv. Opt. Mater. 5, 1700404 (2017).